# Orbital magnetic moments in $FeCr_2S_4$ studied by x-ray magnetic circular dichroism

V. K. Verma[a,b,*], J. Patra[b], V. R. Singh[a,c], Y. Nonaka[a], G. Shibata[a,d], K. Ishigami[a,e], A. Tanaka[f], K. Ohgushi[g], Y. Tokura[h], T. Koide[i], A. Fujimori[a,j]

[a] *Department of Physics, University of Tokyo, Bunkyo-ku, Tokyo 113-0033, Japan*
[b] *Department of Physics, School of Advanced Sciences, VIT-AP University, Amaravati 522241, A.P., India*
[c] *Department of Physics, Central University of South Bihar, Gaya 824236, Bihar, India*
[d] *Materials Sciences Research Center, Japan Atomic Energy Agency, Sayo 679-5148, Japan*
[e] *Japan Synchrotron Radiation Research Institute, Sayo 679-5198, Japan*
[f] *Graduate School of Advanced Science and Engineering, Hiroshima University, Higashi-hiroshima, Hiroshima 739-8530, Japan*
[g] *Department of Physics, Graduate School of Science, Tohoku University, 6-3 Aramaki-Aoba, Aoba-ku, Sendai, Miyagi 980-8578, Japan*
[h] *RIKEN Center for Emergent Matter Science (CEMS), Wako, 351-0198, Japan*
[i] *Photon Factory, IMSS, High Energy Accelerator Research Organization, Tsukuba, Ibaraki 305-0801, Japan*
[j] *Department of Physics and Center for Quantum Science and Technology, National Tsing Hua University, Hsinchu 300044, Taiwan*

[*] Corresponding author: virendra.verma@vitap.ac.in

## Abstract

We have investigated the element specific magnetic characteristics of single-crystal $FeCr_2S_4$ using x-ray absorption spectroscopy (XAS) and x-ray magnetic circular dichroism (XMCD). We have found that the Fe $L_{2,3}$-edge XAS spectra do not exhibit clear multiplet structures, indicating strong hybridization between the Fe 3$d$ and S 3$p$ orbitals, leading to delocalized rather than localized electronic states. The Fe 3$d$ and Cr 3$d$ spin moments are antiferromagnetically coupled, consistent with the Goodenough-Kanamori rule. The orbital magnetic moments of Fe and Cr are determined to be -0.23 and -0.017 $\mu_B$/ion, respectively. The large orbital magnetic moment of Fe is due to the $d^6$ configuration under the relatively weak tetrahedra crystal field at the Fe site, and the delocalized Fe electrons maintain the orbital degree of freedom in spite of their itinerant nature. To understand phenomena such as the gigantic Kerr rotation, it is essential to consider not only the orbital degrees of freedom but also the role of spin-orbit coupling, which induces a finite orbital magnetic moment through $t_2$ and $e$ level hybridization under the tetrahedral crystal field. This finite orbital moment serves as a direct indicator of spin-orbit interaction strength and links element-specific orbital magnetism to the large Kerr rotation. On the other hand, the octahedral crystal-field splitting of the Cr 3$d$ level is large enough to result in the quenching of the orbital moment of the Cr ion in $FeCr_2S_4$.

## Introduction

Transition-metal sulfides with the spinel structure $AB_2X_4$ have attracted attention in both theoretical as well as experimental points of view for to their unique physical properties [1-3]. In $FeCr_2S_4$, the $Fe^{2+}$ ion occupies the tetrahedral site ($A$ site) with the electron configuration of $e^3t_{2g}^3$, and the $Cr^{3+}$ ion occupies the octahedral site ($B$ site) with the electronic configuration of $t_{2g}^3$. The $Fe^{2+}$ ion in $FeCr_2S_4$ is Jahn-Teller-active. Below 9 K, it exhibits an orbital order, which reflects the $e$-orbital degree of freedom of the $Fe^{2+}$ ion [4-7]. On the other hand, the $Cr^{3+}$ ion is Jahn-Teller-inactive because of the half-filled $t_{2g}$ level. The $Fe^{2+}$ ion ($3d^6$ configuration) at the tetrahedral site, can retain the orbital degree of freedom and sizeable orbital magnetic moment under a relatively weak crystal field splitting. This feature makes $FeCr_2S_4$ a candidate system for exploring orbital-driven physics in correlated materials. Within the last decade, several novel physical phenomena were discovered in these compounds such as spin-orbital-liquid behavior [8], colossal magneto-resistance [9,10,11], orbital glasses [12], colossal magneto-capacitive coupling [13], and complex spin order and spin dimerization [14,15]. These novel phenomena in transition metal-based spinel sulfides have piqued interest in spintronic and advanced multifunctional device applications.

$FeCr_2S_4$ is identified as a ferromagnetic semiconductor exhibiting the Curie temperature of approximately 170 K and significant negative magnetoresistance around this temperature [10]. Strong coupling among the spin, orbital, charge, and lattice degrees of freedom in $FeCr_2S_4$ results in unusual low-field magnetic behavior [16] and related fascinating physical properties such as gigantic Kerr rotation [17] and colossal magnetoresistance [10,11]. For these remarkable physical properties, spin-orbit interaction and orbital magnetic moments are believed to play important roles. Park *et al*. [18] studied the orbital magnetic moment by band-structure calculation within density-functional theory (DFT). They found that the sublattices of Fe and Cr are antiferromagnetically coupled to each other, while each sublattice of Fe and Cr is ferromagnetically ordered. Recently, band structure calculation by Sarkar *et al*. [19] using DFT+U found that the Cr ion has a small orbital moment while the Fe ion has a large orbital moment. In recent years, several theoretical and experimental studies on this compound have been conducted but the direct experimental evidence of the orbital contribution in this material has been lacking. Hence, the origin and the roles of the enhanced orbital magnetic moment of Fe in $FeCr_2S_4$ still require further investigations through element specific spectroscopic techniques. The orbital degrees of freedom, which arises from the minority-spin electron in the $e$ level of the $Fe^{2+}$ ion, plays an important role in the interesting physical properties of $FeCr_2S_4$.

However, to understand interesting phenomena such as the gigantic Kerr rotation [17], spin-orbital liquid behavior [8], and non-collinear spin structures at low temperatures [16], not only the orbital degrees of freedom but also substantial spin-orbit coupling and induced orbital magnetic moment are necessary. The orbital magnetic moment of the minority-spin electron in the *e* level of the tetrahedrally co-ordinated $Fe^{2+}$ ion, unlike the $t_2$ level, should be quenched by the $T_d$ crystal field in the absence of spin-orbit coupling, but will be partially recovered under finite spin-orbit coupling, which causes hybridization between the $t_2$ and *e* levels and induces orbital moment in the *e* levels. In order to confirm this scenario and to study how large orbital magnetic moment is induced in the $Fe^{2+}$ ion, we have carried out the element-specific x-ray magnetic circular dichroism (XMCD) technique which provides direct insights into the electronic structure, the spin and orbital magnetic moments of the Fe and Cr ions in $FeCr_2S_4$. Here, we report on an investigation of the electronic and magnetic properties of the Fe and Cr ions in $FeCr_2S_4$ single crystal by x-ray absorption spectroscopy (XAS) and XMCD spectroscopic measurements at the Fe and Cr $L_{2,3}$ edges, and study the microscopic origin of the large orbital magnetic moment of Fe in $FeCr_2S_4$.

## Experimental

$FeCr_2S_4$ single crystals were grown using the chemical vapour transport technique, employing the transport agent of $CrCl_3$. The ground crystals were characterized by powder x-ray diffraction, which confirmed that the obtained crystals were single phase [17]. In order to obtain a clean surface, the sample was cleaved *in-situ* in the measurement chamber. The XAS and XMCD measurements at the Fe and Cr $L_{2,3}$ edges were done at the undulator beamline BL-16 of Photon Factory (PF), KEK. Spectroscopic data were collected in the total-electron yield (TEY) mode (probing depth ～5 nm) at $T$ = 80 K by measuring the sample current and normalizing it to the mirror current. XMCD spectra were obtained by reversing the direction of photon helicity at a fixed magnetic field. Magnetic field were applied using a superconducting magnet. The directions of the x-rays and the applied magnetic field with respect to the $FeCr_2S_4$ crystal are shown in Fig. 1. The base pressure of the measurement chamber was maintained at about $10^{-9}$ Torr and the energy resolution was $E/\Delta E \sim 8{,}000$.

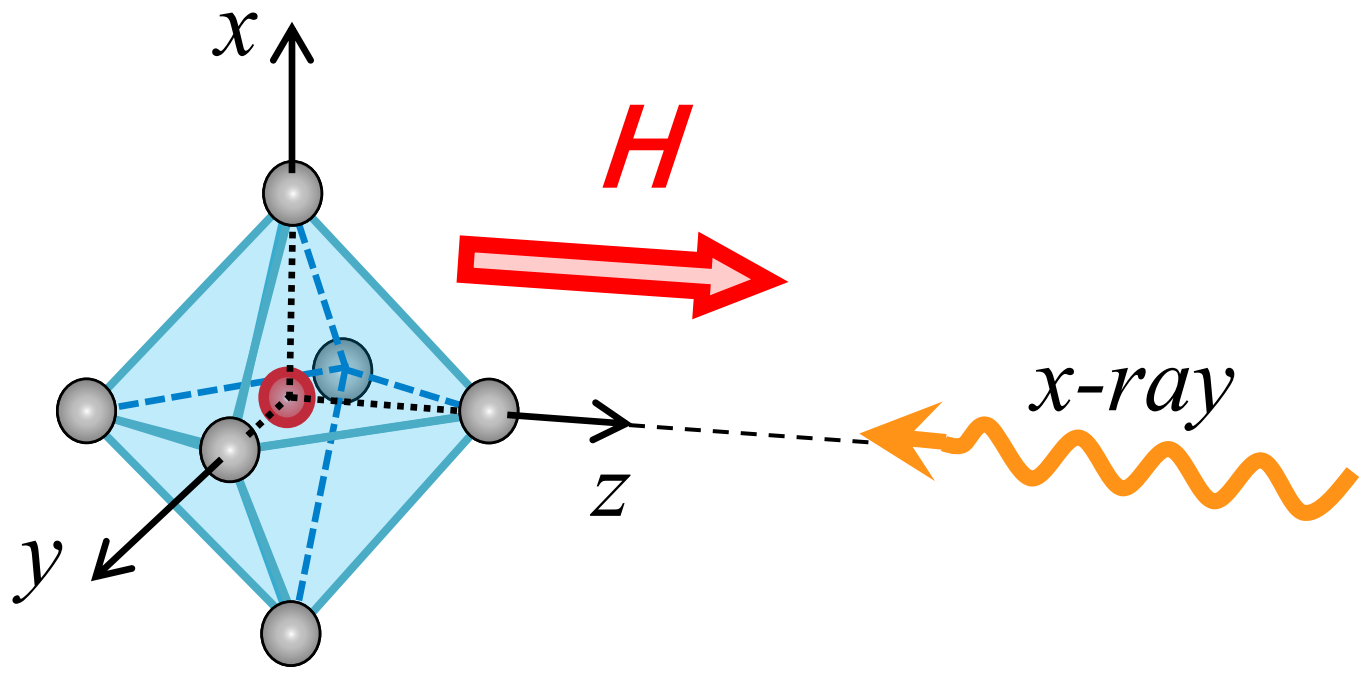

Figure 1: Directions of the x ray and the applied magnetic field with respect to the crystal axes of the single-crystalline $FeCr_2S_4$ sample.

## Results and Discussion

Figure 2(a) shows the Fe $L_{2,3}$-edge XAS spectra of $FeCr_2S_4$ under a magnetic field of 1 T for incident x rays with both positive and negative helicities ($\mu_+$ and $\mu_-$) relative to the Fe spin taken at 80 K. The major two peaks around $h\nu$ = 707.9 and 720.6 eV correspond to absorption from the Fe $2p_{3/2}$ and $2p_{1/2}$ core levels to unoccupied Fe 3*d* states, respectively. In the $L_3$-edge ($2p_{3/2}$ core-absorption) region, the XAS intensity for helicity $\mu_+$ is larger than that of $\mu_-$, whereas in the $L_2$-edge ($2p_{1/2}$ core absorption) region, the XAS intensity for helicity $\mu_+$ is smaller than that of $\mu_$. The difference between the $\mu_+$ and $\mu_-$ XAS derives the XMCD ($\mu_+$-$\mu_-$) spectra. The Fe $L_{2,3}$-edge XMCD spectra measured at different magnetic fields are shown in Fig. 2 (b). The magnified Fe $L_{2,3}$-edge XMCD taken at 0 T is shown in the inset. The non-zero intensity of XMCD at $H$ = 0 T, which was taken after applying $H$ = 0.1 T, clearly indicates a remanent magnetization involving the Fe ions in the sample.

Figure 2 (c) shows the Fe $L_{2,3}$-edge XAS spectrum of $FeCr_2S_4$ plotted with those of FeO ($Fe^{2+}$) [20], Fe metal [21], and γ-$Fe_2O_3$ ($Fe^{3+}$) [22]. By comparing all the Fe $L_{2,3}$-edge XAS spectra, it is clearly seen that the line shape of Fe $L_{2,3}$-edge XAS spectrum of $FeCr_2S_4$ are similar to that of FeO but different from that of γ-$Fe_2O_3$. However, unlike FeO, no (or very weak) multiplet features are seen at the $L_3$ and $L_2$ edges of $FeCr_2S_4$, resulting in a line shape that is remarkably close to that of Fe metal. The absence of multiplet features in the Fe $L_{2,3}$-edge XAS of $FeCr_2S_4$ indicates a substantial hybridization between the Fe 3*d* orbitals and the S 3*p* valence orbitals. Nevertheless, the Fe $L_{2,3}$-edge XAS spectra of FeO and $FeCr_2S_4$ are narrower than the spectrum of Fe metal. The strong hybridization between Fe 3*d* and S 3*p* orbitals, the valence electrons become more delocalized, and the multiplet splitting is significantly obscured or even suppressed, resulting in smoother, featureless XAS line shapes. This is particularly true for the $Fe^{2+}$ ion in the tetrahedral coordination with sulfur, as in $FeCr_2S_4$, where Fe 3*d* orbitals strongly overlap with the more diffuse and less electronegative S 3*p* orbitals. Such hybridization increases the covalency of the Fe-S bond and causes screening of the 2*p* core-hole, which reduces the multiplet splitting observed in XAS. Figure 2 (d) shows the Fe 2*p* XMCD spectra of $FeCr_2S_4$ and Fe metal [21]. Contrary to the XAS spectra, the Fe $L_{2,3}$-edge XMCD spectrum of $FeCr_2S_4$ shows some multiplet features indicated by arrows while the multiplet features are absent in the $L_{2,3}$-edge XMCD of Fe metal. From Figs. 2 (c) and (d), one can conclude that the Fe ions in $FeCr_2S_4$ are mainly in the divalent (2+) state. The valence $Fe^{2+}$ in $FeCr_2S_4$ has also been confirmed by Mossbauer spectroscopy measurements performed by Chen *et al*. [23] and Pyataev *et al*. [24].

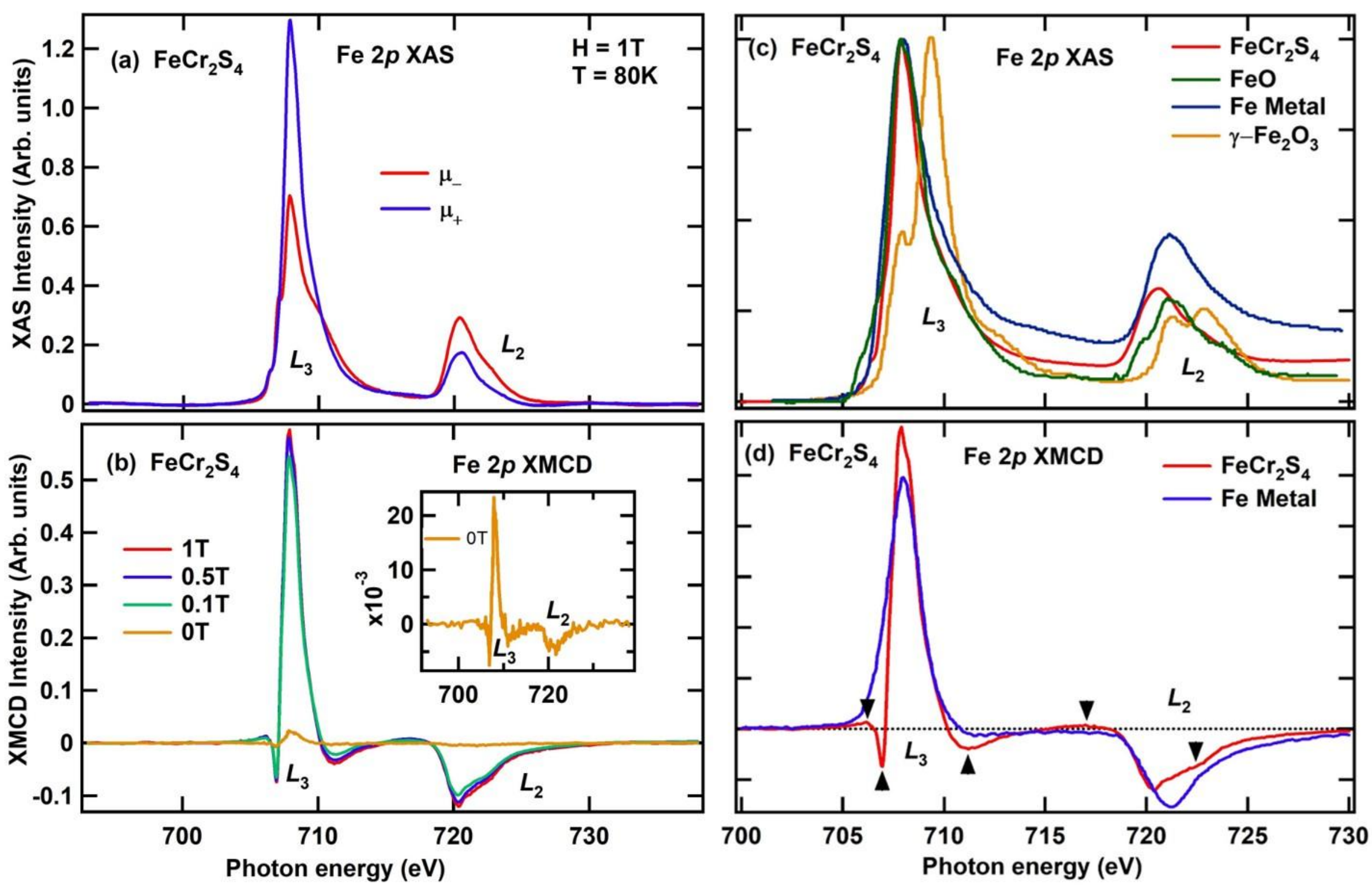

Figure 2: Fe $L_{2,3}$-edge XAS and XMCD spectra of $FeCr_2S_4$. (a) Fe $L_{2,3}$-edge XAS spectra under a magnetic field of 1 T for incident x rays with both positive and negative helicities ($\mu_+$ and $\mu_-$). (b) Fe $L_{2,3}$-edge XMCD spectra at various magnetic fields. The inset shows a magnified plot at 0 T, which was taken after applying $H$ = 0.1 T, indicating that there was a remanent magnetization involving Fe. (c) Comparison of the Fe $L_{2,3}$-edge XAS spectrum of $FeCr_2S_4$ with those of FeO ($Fe^{2+}$) [20], Fe-metal [21] and γ-$Fe_2O_3$ ($Fe^{3+}$) [22]. While $FeCr_2S_4$ shows extremely weak multiplet features while Fe metal shows no multiplet features. (d) XMCD spectrum of $FeCr_2S_4$ compared with that of Fe metal [21]. Some multiplet features are discernible for $FeCr_2S_4$ but not for Fe metal.

Figure 3 (a) shows the Cr $L_{2,3}$-edge XAS spectra of $FeCr_2S_4$ for incident x rays with different helicities ($\mu_+$ and $\mu_-$), i.e., parallel and antiparallel to the Cr 3*d* spin, respectively, taken at 80 K. X-ray absorption from the Cr $2p_{3/2}$ and Cr $2p_{1/2}$ core levels result in two major peaks in the $L_{2,3}$-edge XAS around $h\nu = 579.6$ and 587.9 eV, respectively. It is found that the XAS spectrum for positive helicity ($\mu_+$) is broader than that for negative helicity ($\mu_-$). Furthermore, the $\mu_+$ spectrum is shifted to higher energies in comparison to $\mu_-$ in both the $L_2$ and $L_3$ regions, leading to the complicated Cr $L_{2,3}$-edge XMCD ($\mu_+$-$\mu_-$) line shapes, as shown in Fig. 3 (b). A magnified Cr $L_{2,3}$ XMCD at 0 T is shown in the inset. The non-zero intensity of Cr $L_{2,3}$ XMCD at $H = 0$ T clearly indicates the remanent magnetization involving the Cr ions in $FeCr_2S_4$. From Figs. 2 and 3, one can see that the signs of the Fe and Cr $L_{2,3}$ XMCD are opposite to each other, which indicates that the spin moments of the Fe and Cr ions in $FeCr_2S_4$ are antiparallel to each other, in agreement with the DFT calculations by Park *et al*. [18] and Sarkar *et al*. [19]. Here, the antiferromagnetic alignment of the Fe and Cr ions rules out the simple double-exchange mechanism of electron transfer between the Fe and Cr ions [23]. In Fig. 3 (c), we compare the Cr $L_{2,3}$-edge XAS spectra of $FeCr_2S_4$, $Cr_2O_3$ ($Cr^{3+}$) [25], $Zn_{1-x}Cr_xTe$ [26] and Cr metal [27]. From the figure, it is evident that the Cr $L_{2,3}$ XAS spectrum of $FeCr_2S_4$ exhibits significant differences from those of $Zn_{1-x}Cr_xTe$, where Cr is in the 2+ valence state, and Cr metal, which shows spin-density waves (SDW) [28]. This SDW ground state leads to broad and featureless XAS line shapes, reflecting the delocalized nature of the Cr 3*d* electrons and the lack of crystal-field-induced multiplet splitting. On the other hand, it shows a qualitative similarity to the spectrum of $Cr_2O_3$, thereby confirming that the Cr ions in $FeCr_2S_4$ predominantly exist in the trivalent (3+) state and hence have the high-spin $t_{2g}^3$ configuration. The Cr *L*-edge XAS spectrum of $FeCr_2S_4$ displays sharper and more structured features, which are characteristic of localized $Cr^{3+}$ ions in an octahedral ligand field. However, in contrast to the XAS of $Cr_2O_3$, the multiplet features in $FeCr_2S_4$ are less pronounced. The observed differences in the multiplet features indicate that the Cr-S bond is more covalent than the Cr-O bond in $Cr_2O_3$ due to the lower electronegativity of S than O and that the Cr 3*d* electrons are more delocalized in $FeCr_2S_4$ than those in $Cr_2O_3$.

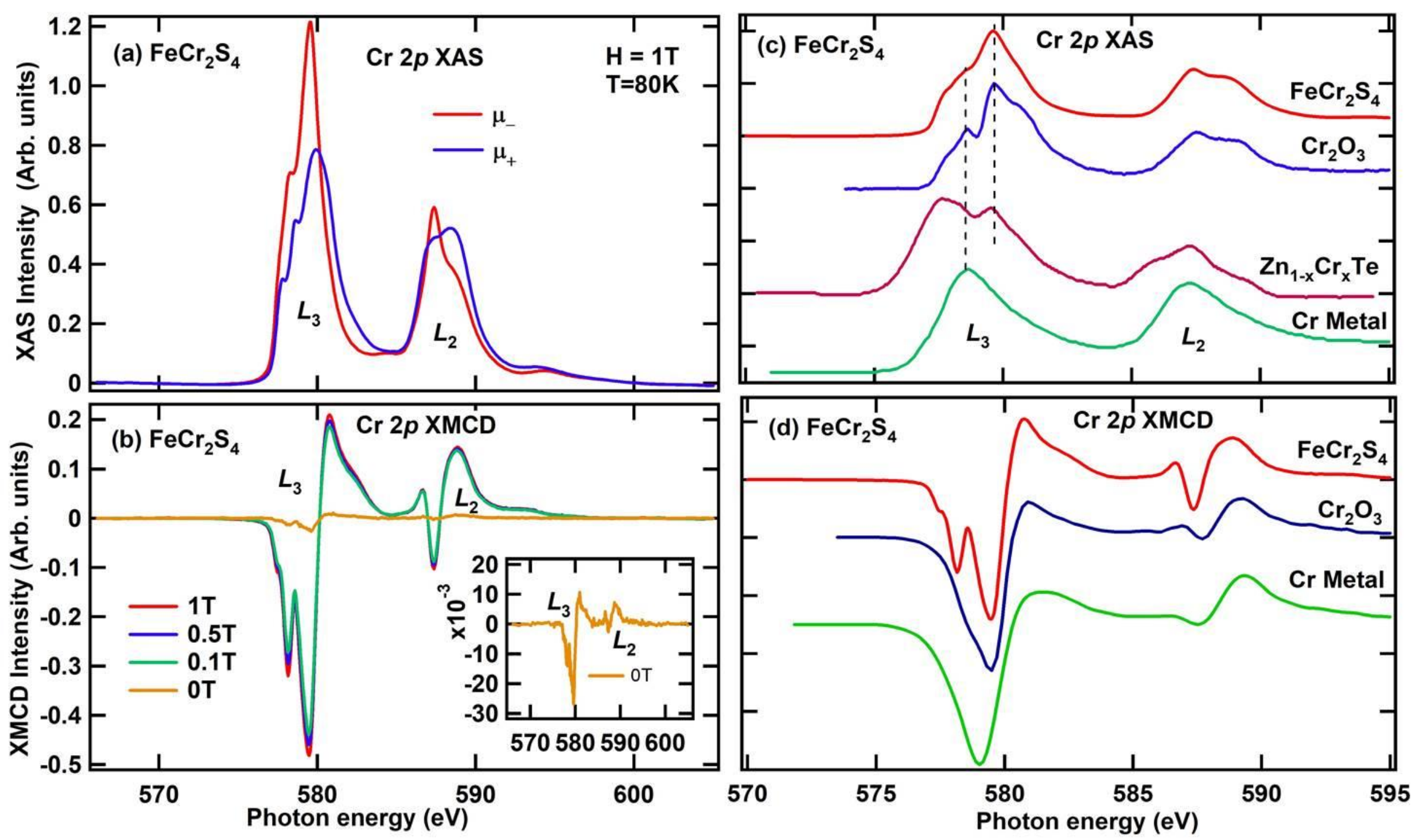

Figure 3: Cr $L_{2,3}$-edge XAS and XMCD spectra of $FeCr_2S_4$. (a) Cr $L_{2,3}$-edge XAS spectra under a magnetic field of 1 T for incident x rays with opposite helicities. (b) Cr $L_{2,3}$-edge XMCD spectra at various magnetic fields. The inset shows a magnified plot at 0 T, indicating that there was a remnant magnetization involving Cr. (c) Comparison of the Cr $L_{2,3}$-edge XAS spectrum of $FeCr_2S_4$ with those of $Cr_2O_3$ ($Cr^{3+}$) [25], $Zn_{1-x}Cr_xTe$ ($Cr^{2+}$) [26], and Cr metal [27]. (d) XMCD spectrum of $FeCr_2S_4$ compared with those of $Cr_2O_3$ [25] and Cr metal [27].

We have evaluated the spin and orbital magnetic moments ($M_{\mathrm{spin}}$ and $M_{\mathrm{orb}}$) of the Fe and Cr ions using the XMCD sum rules [21,29,30]

$$M_{\mathrm{spin}} + 7m_T = -\frac{6\int_{L_3}(\mu^+ - \mu^-)d\omega - 4\int_{L_3+L_2}(\mu^+ - \mu^-)d\omega}{\int_{L_3+L_2}(\mu^+ + \mu^-)d\omega} N_d \quad , \quad (1)$$

$$M_{\mathrm{orb}} = -\frac{4\int_{L_3+L_2}(\mu^+ - \mu^-)d\omega}{3\int_{L_3+L_2}(\mu^+ + \mu^-)d\omega} N_d \quad , \quad (2)$$

The $M_{\mathrm{spin}}$ and $M_{\mathrm{orb}}$ are given in units of Bohr magneton ($\mu_B$) per atom. $N_d$ is the number of unoccupied $3d$ electrons, and $m_T$ is the magnetic dipole moment. The $m_T$ is negligible compared to $M_{\mathrm{spin}}$ and is omitted from the calculations [31]. For the calculation of magnetic moments by the XMCD sum rules, we adopted $N_d = 4$ for $Fe^{2+}$ ($3d^6$) and $N_d = 7$ for $Cr^{3+}$ ($3d^3$). The Fe $L_{2,3}$-edge XAS and XMCD spectra and their energy integrals are shown in Figs. 4(a) and 4(b), and the Cr $L_{2,3}$-edge XAS and XMCD spectra with their energy integrals are shown in Figs. 4(c) and 4(d). The black dotted lines in Figs. 4 (a) and 4 (c) show an arctangent background to be subtracted [21]. The spin and orbital magnetic moments of Fe and Cr in $FeCr_2S_4$ evaluated using the XMCD sum rules are listed in Table 1, together with those obtained from the cluster-model analyses (see below) and the DFT calculation [19]. Note that, for light transition-metal elements like Cr, precise estimation of spin magnetic moment by using XMCD sum rules is hindered by the overlap between the $L_2$ and $L_3$ edges.

Table 1. Magnetic moments of Fe and Cr in $FeCr_2S_4$ obtained from experiment using the XMCD sum rules, cluster-model analyses and DFT calculations.

| Elements in $FeCr_2S_4$ | Fe | | Cr | |
|---|---|---|---|---|
| Magnetic moment ($M$) | $M_{spin}$ ($\mu_B$ /ion) | $M_{orb}$ ($\mu_B$ /ion) | $M_{spin}$ ($\mu_B$ /ion) | $M_{orb}$ ($\mu_B$ /ion) |
| Experiment: XMCD sum rules | -1.99 | -0.23 | 1.48 | -0.017 |
| Cluster-model calculation | -1.55 | -0.15 | 1.15 | -0.024 |
| DFT (GGA+SO)[#] [19] | -3.13 | -0.08 | 2.75 | -0.024 |
| DFT (GGA+U+SO)[#] [19] | -3.27 | -0.13 | 2.69 | -0.026 |

[#]GGA, $U$, and SO stands for generalized gradient approximation, Hubbard $U$, and spin-orbit interaction, respectively.

The opposite signs of the evaluated spin magnetic moments between Fe and Cr follow the Goodenough-Kanamori rule [32]. The evaluated values of the orbital magnetic moment for the Fe and Cr ions by the sum rule analysis were -0.23 and -0.017 $\mu_B$, respectively. The evaluated value of the spin magnetic moment for the Fe ion was -1.99 $\mu_B$. However, strong hybridization between Fe 3$d$ and S 3$p$ orbitals leads to partial covalent character in the Fe-S bond and charge delocalization, where some electron density from the ligand (S 3$p$) orbitals is transferred into the Fe 3$d$ orbitals. As a result, the 3$d$ occupancy at Fe site increases and hence $N_d$ is less than 4 for $Fe^{2+}$, possibly around 3.2–3.8 (decrease 5 to 20 %) [21,31,33] depending on the hybridization strength and the spin and orbital magnetic moments of Fe reduces from their original values of -1.99 $\mu_B$ and -0.23 $\mu_B$ to the ranges of -1.59 to -1.89 $\mu_B$ and 0.18 to 0.22 $\mu_B$, respectively. Our results indicate that the Fe ion exhibits identical signs for both the spin and orbital magnetic moments, whereas the Cr ion displays opposite signs. The different relative signs of the spin and orbital magnetic moments between the Fe and Cr ions occur due to the fact that the $d$ states of $Fe^{2+}$ ions are more than half filled, whereas those of $Cr^{3+}$ ions are less than half filled, according to Hund's third rule. The Fe ions exhibit a significant orbital moment, which may be attributed to the relatively small crystal-field splitting of the Fe 3$d$ level into the $e$ and $t_2$ levels at the $A$ site of the spinel structure. The relatively strong intra-atomic $d$-$d$ transition ($^5E \rightarrow {}^5T_2$) reported by Ohgushi *et al.* [17] is due to inversion symmetry breaking at the $Fe^{2+}$ site and the stronger covalency in the chalcogenides than the halides and oxides. The strong magneto-optical effect related to the $Fe^{2+}$ ion in $FeCr_2S_4$ can be attributed to the finite orbital magnetic moment, the spin-orbit coupling, and inversion symmetry breaking inherent at the $A$-site, in contrast to the $Mn^{2+}$ ion in $MnCr_2S_4$, where the spin-orbit coupling is virtually absent [17]. In this context, a finite orbital magnetic moment is a direct indicator of spin-orbit interaction strength. Therefore, the XMCD findings provide evidence supporting the origin of the large Kerr rotation, establishing a direct link between the element-specific orbital magnetism and the magneto-optical behavior observed in the earlier MOKE experiments. On the other hand, the $d^3$ configuration of the Cr ion results in a small orbital moment. Therefore, the half-filled $t_{2g}$ shell with $S$=3/2 of $Cr^{3+}$ at the octahedral site behaves like a spin-only system.

Theoretical spectra of the Fe and Cr $L_{2,3}$-edge XAS and XMCD have been derived from cluster-model calculations as shown in Fig. 5. The calculated spin and orbital magnetic moments using the best-fit parameters are listed in the second row of Table 1. The fitting parameters for the Fe $L_{2,3}$ edge were $\Delta$ = 0.5 eV, $U_{dd}$ = 5 eV, $pd\sigma$ = 0.7 eV, and $10Dq$ = 0.4 eV and those for the Cr $L_{2,3}$ edge were $\Delta$ = 1.7 eV, $U_{dd}$ = 2.5 eV, $pd\sigma$ = 1.1 eV and $10Dq$ = 1.5 eV, where $\Delta$ is the charge-transfer energy, $U_{dd}$ is the $d$-$d$ Coulomb repulsion energy, $pd\sigma$ is the transfer energy, and $10Dq$ is the crystal-field splitting energy. The calculated spectra for Fe and Cr $L_{2,3}$-edge XAS and XMCD are consistent with the experimental results including the highly asymmetric line shape of the Fe $L_{2,3}$-edge XAS, although the multiplet structures of the Fe $L_{2,3}$-edge XAS and XMCD are more pronounced than experiment. The orbital and spin magnetic moments calculated using the cluster model with the above parameters are, respectively, -0.15 and -1.55 $\mu_B$/ion for Fe, and -0.024 and 1.15 $\mu_B$/ion for Cr. Sarkar *et al.* [19] reported the orbital moments of Fe and Cr to be -0.13 and -0.026 $\mu_B$/ion, respectively, from the DFT calculation. As indicated in Table 1, the orbital moment of the Fe ion obtained from our experimental result is a little larger than the DFT (GGA+SA) calculation, whereas the orbital magnetic moment of Cr ion is somewhat smaller than the value reported by DFT+U (GGA+SA+U) calculation [19]. The Fe 3$d$-S 3$p$ transfer integral ($pd\sigma$) = 0.7 eV for Fe in $FeCr_2S_4$ is much smaller than the $pd\sigma$ of FeO. This smaller $pd\sigma$ value may imply smaller orbital overlap between Fe 3$d$ and S 3$p$ orbitals, which would decrease the covalent character of the Fe-S bond and makes the Fe 3$d$–S 3$p$ hybridization weaker. However, the charge transfer energy $\Delta$ = 0.5 eV in $FeCr_2S_4$ is much smaller than $\Delta$ = 6 eV in FeO [34], increasing the $p$-$d$ hybridization. The later effect overwhelms the smaller ($pd\sigma$) and increases the covalency in $FeCr_2S_4$.

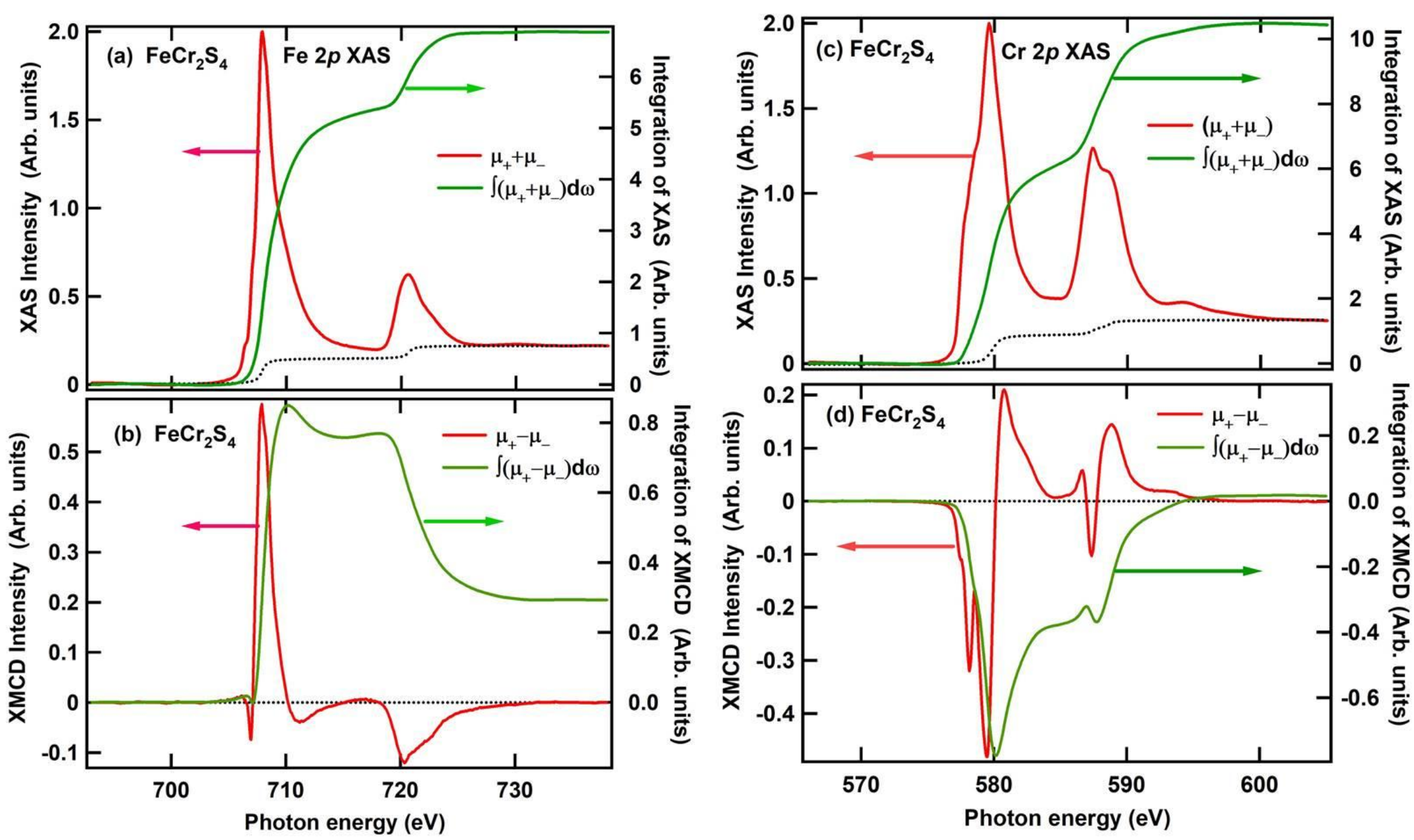

Figure 4: Sum-rule analysis of the XAS and XMCD spectra of $FeCr_2S_4$. (a) Fe 2*p* ($\mu_+$+$\mu_-$) XAS spectrum and its energy integral. (b) XMCD ($\mu_+$-$\mu_-$) spectrum and its energy integral. (c) Cr 2*p* ($\mu_+$+$\mu_-$) XAS spectrum and its energy integral. (d) XMCD ($\mu_+$-$\mu_-$) spectrum and its energy integral.

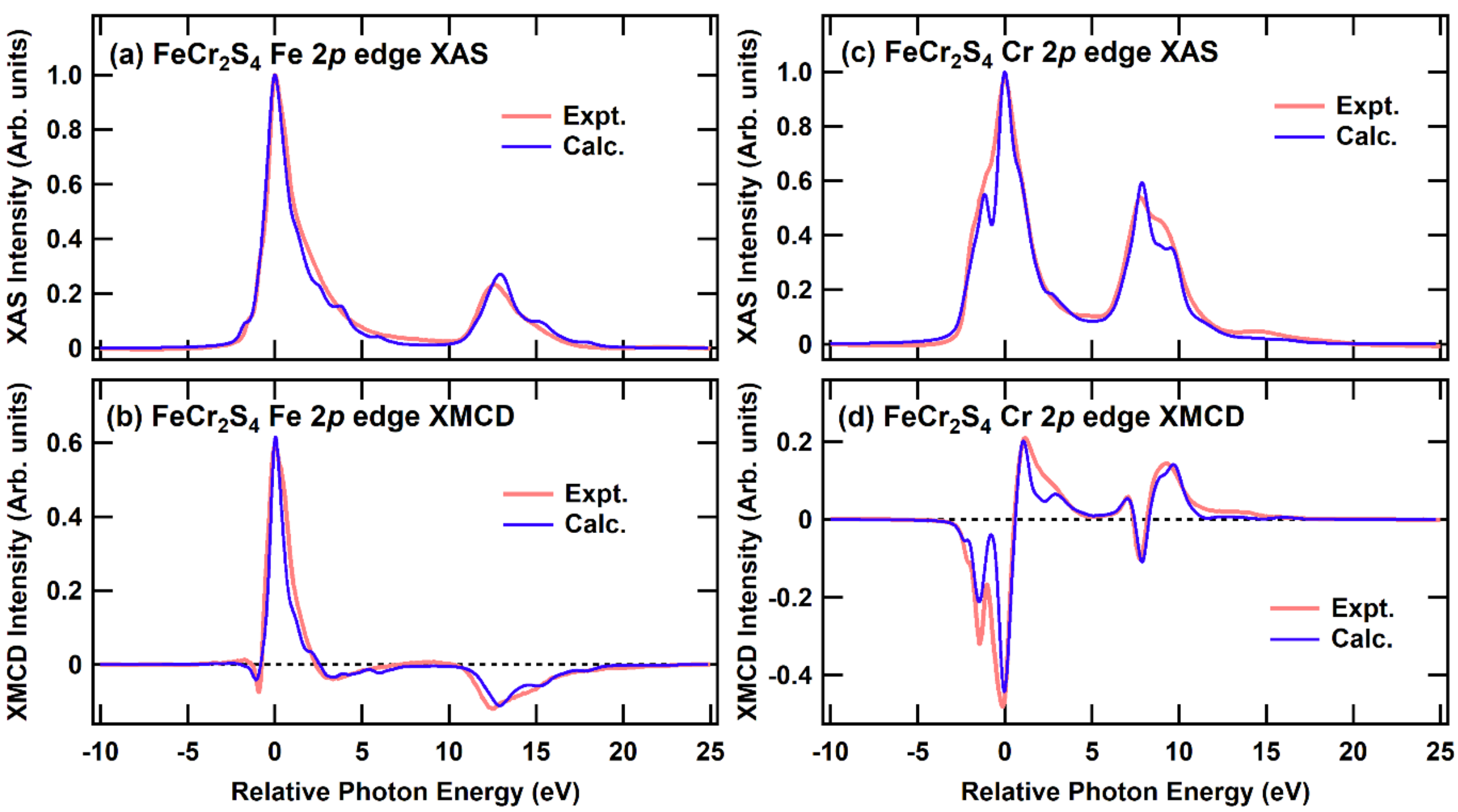

Figure 5. Calculated spectra for Fe and Cr $L_{2,3}$-edge XAS and XMCD using the cluster model compared with the experimental results.

The XAS and XMCD measurements of $Fe_{0.5}Cu_{0.5}Cr_2S_4$ performed by Han *et al.* [35]

reported the parallel alignment of spin and orbital magnetic moments of $Cr^{3+}$. Hund's third rule states that for systems with more than a half-filled *d*-orbital, the spin and orbital moments should align parallel while, for those with less than a half-filled *d*-orbital, they should align antiparallel. The strong octahedral crystal-field splitting in the $Cr^{3+}$ ion ($d^3$ system) leads to the half-filled $t_{2g}$ shell ($t_{2g}^3$, S = 3/2) with no orbital degeneracy, resulting in an almost fully quenched orbital magnetic moment. However, spin-orbital coupling between the crystal-field-split $t_{2g}$ and $e_g$ level induces a small orbital magnetic moment that is aligned anti-parallel to the spin magnetic moment according to Hund's third rule. However, we note that XMCD analysis can be challenging, particularly for small orbital moments. Nevertheless, the orbital moment obtained from our cluster-model calculations, as well as from DFT calculations by Sarkar *et al.* [19], is in qualitative agreement with the experimental value, strengthening the validity of our findings. Park *et al.* [18] and Sarkar *et al.* [19] also found that the opposite alignment of the spin and orbital moments of Cr ions from band structure calculations for $FeCr_2S_4$, consistent with our experimental findings.

## Conclusion

$FeCr_2S_4$ single crystals grown using the chemical vapour transport technique were studied by XAS and XMCD. The line-shape analysis of XAS and XMCD determined that the valence states of the Fe and Cr ions are almost divalent ($Fe^{2+}$) and trivalent ($Cr^{3+}$), respectively. The Fe $L_{2,3}$-edge spectra exhibit very weak multiplet features, which suggests strong hybridization between the Fe 3*d* and S 3*p* electrons. From the magnetic-field dependence of the XMCD measurement of the Fe and Cr ions, we conclude that the Fe and Cr ions ordered ferromagnetically in each sublattice. From the sum rule analysis, the Fe and Cr sublattices are coupled antiferromagnetically to each other. The spin and orbital magnetic moments of Cr are aligned in the opposite directions, in agreement with Hund's third rule. A large orbital moment of Fe (-0.23 $\mu_B$/ion) was deduced by a sum rule analysis of the element specific XMCD data. Our findings support the scenario that the orbital magnetic moment in $Fe^{2+}$ arises due to spin-orbit coupling, which partially restores the otherwise quenched orbital moment in the *e* level through hybridization with the $t_2$ level. This induced orbital moment not only serves as a direct indicator of spin-orbit interaction strength but also links element-specific orbital magnetism to the large Kerr rotation observed in $FeCr_2S_4$. On the other hand, the orbital magnetic moment of Cr is almost quenched due to the half-filled $t_{2g}$ shell.

## Acknowledgement

The authors thank Tanusri Saha-Dasgupta and Indra Dasgupta for valuable discussion and Kenta Amemiya and Masako Sakamaki for their valuable technical support at the Photon Factory. The experiment was performed under the approval of the program Advisory Committee (Proposal Nos. 2011G582, 2010S2-001). VKV would like to thank University Grant Commission, Department of Atomic Energy, Consortium for Scientific Research (UGC-DAE CSR) (CRS/2022-23/01/726) for their financial support. AF would like to thank the Japan Society for the Promotion of Science (KAKENHI grant No. JP-22K03535), the National Science and Technology Council of Taiwan (grant No. NSTC113-2112-M-007-033), and the Yushan Fellow Program and the Center for Quantum Science and Technology within the framework of the Higher Education Sprout Project under the Ministry of Education of Taiwan.

## APPENDIX

The $FeCr_2S_4$ single crystals were gently crushed into fine powder and characterized using powder X-ray diffraction (XRD), as shown in Fig. A1. The XRD pattern exhibits well-defined diffraction peaks, confirming that the obtained crystals are single-phase and possess the expected chemical composition without any detectable secondary phases or impurities.

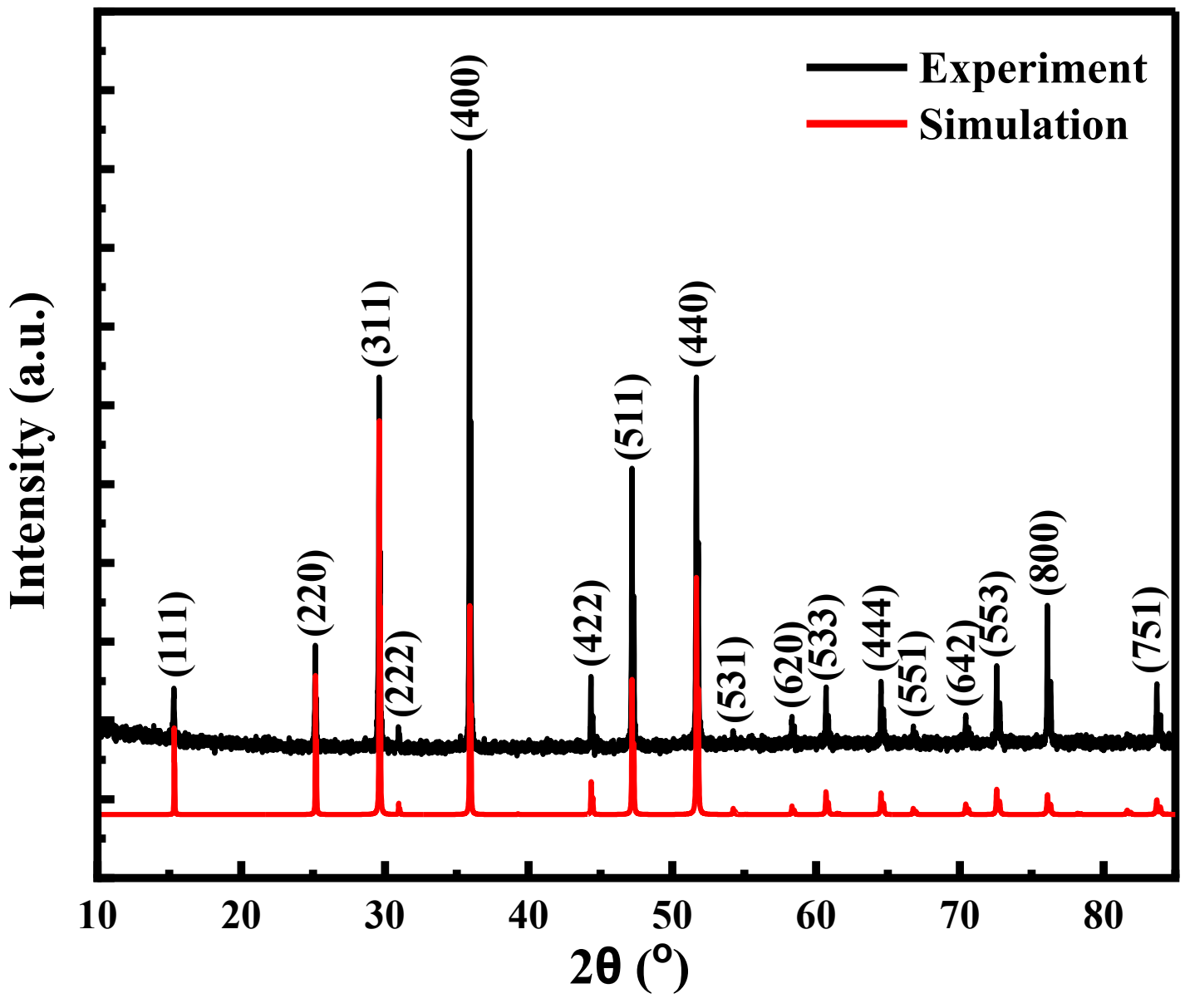

Figure A1: XRD Pattern of crushed single crystal $FeCr_2S_4$

## References

[1] A. Kimura, J. Matsuno, J. Okabayashi, and A. Fujimori, T. Shishidou, E. Kulatov, T. Kanomata, Phys. Rev. B **63**, 224420 (2001).

[2] P.G. Radaelli, New J. Phys. **7**, 53 (2005).

[3] K. Ohgushi, Y. Okimoto, T. Ogasawara, S. Miyasakai, and Y. Tokura, J. of the Phys. Soc. of Japan **77**, 034713 (2008).

[4] L. Prodan, S. Yasin, A. Jesche, J. Deisenhofer, H.-A. Krug von Nidda, F. Mayr, S. Zherlitsyn, J. Wosnitza, A. Loidl, V. Tsurkan, Phys. Rev. B **104**, L020410 (2021).

[5] F.K. Lotgering, A.M. Van Diepen, J.F. Olijhoek, Solid State Commun. **17**, 1149 (1975).

[6] L. Brossard, J.L. Dormann, L. Goldstein, P. Gibart, P. Renaudin, Phys. Rev. B **20**, 2933 (1979).

[7] L.F. Feiner, J. Phys. C **15**, 1515 (1982).

[8] V. Fritsch, J. Hemberger, N. Buttgen, E.-W. Scheidt, H.-A. Krug Von Nidda, A. Loidl, V. Tsurkan, Phys. Rev. Lett. **92**, 116401 (2004).

[9] N. Jia, B. Fang and Z. Yang, Appl. Phys. A **116**, 839 (2014).

[10] A.P. Ramirez, R.J. Cava, and J. Krajewski, Nature **386**, 156 (1997).

[11] L. Liu, Q. Yu, J. Xia, W. Shi, D. Wang, J. Wu, L. Xie, Y. Chen, and L. Jiao, Adv. Mater. **36**, 2401338 (2024).

[12] C. Gu, Z. Yang, R. Tong, X. Chen, Y. Sun, L. Pi, and Y. Zhang, J. Phys.: Condens. Matter **27**, 026003 (2015).

[13] J. Hemberger, P. Lunkenheimer, R. Fichtl1, H.-A. Krug von Nidda, V. Tsurkan and A. Loidl, Nature **434**, 364 (2005).

[14] P.G. Radaelli, Y. Horibe, M. Gutmann, H. Ishibashi, C. Chen, R. Ibberson, Y. Koyama, Y.-S. Hor, V. Kiryukhin, S.-W. Cheong, Nature **416**, 155 (2002).

[15] J. Bertinshaw, C. Ulrich, A.Gunther, F. Schrettle, M. Wohlauer, S. Krohns, M. Reehuis, A.J. Studer, M. Avdeev, D.V. Quach, J.R. Groza, V. Tsurkan, A. Loidl & J. Deisenhofer, Sci. Rep. **4**, 6079 (2014).

[16] V. Tsurkan, V. Fritsch, J. Hemberger, H.-A. Krug von Nidda, N. Buttgen, D. Samusi, S. Ko¨rner, E.-W. Scheidt, S. Horn, R. Tidecks, A. Loidl, J. Phys. Chem. Solids, **66**, 2036 (2005).

[17] K. Ohgushi, T. Ogasawara, Y. Okimoto, S. Miyasaka, and Y. Tokura, Phys. Rev. B **72**, 155114 (2005).

[18] M.S. Park, S.K. Kwon, S.J. Youn, and B.I. Min, Phys. Rev. B **59**, 10018 (1999).

[19] S. Sarkar, M.D. Raychaudhury, I. Dasgupta, and T.S. Dasgupta, Phys. Rev. B **80**, 201101(R) (2009).

[20] T.J. Regan, H. Ohldag, C. Stamm, F. Nolting, J. Luning, and J. Stohr, R.L. White, Phys. Rev. B **64**, 214422 (2001).

[21] C.T. Chen, Y.U. Idzerda, H.-J. Lin, N.V. Smith, G. Meigs, E. Chaban, G.H. Ho, E. Pellegrin, and F. Sette, Phys. Rev. Lett. **75**, 152 (1995).

[22] S.B. Profeta, M.-A. Arrio, E. Tronc, N. Menguy, I. Letard, C.C. D. Moulin, M. Nogues, C. Chaneac, J.-P. Jolivet and P. Sainctavit, J. Magn. and Magn. Mater. **288**, 354 (2005).

[23] Z. Chen, S. Tan, Z. Yang, and Y. Zhang, Phys. Rev. B **59**, 11172 (1999).

[24] A. Pyataev, I. Malikov, E. Voronina, and E. Dulov, J. Mol. Struct. **1199**, 126941 (2020).

[25] C. Theil, J. van Elp, and F. Folkmann, Phys. Rev. B **59**, 7931 (1999).

[26] Y. Yamazaki, T. Kataoka, V.R. Singh, A. Fujimori, F.-H Chang, D.-J Huang, H.-J Lin, C.T. Chen, K. Ishikawa, K. Zhang and S. Kuroda, J. Phys.: Condens. Matter **23** 176002 (2011).

[27] M.A. Tomaz and W.J. Antel, W.L. O'Brien, G.R. Harp, Phys. Rev. B **55**, 3716 (1997).

[28] E. Fawcett, Rev. Mod. Phys. **60**, 209 (1988).

[29] B.T. Thole, P.Carra, F. Sette, and G. van der Laan, Phys. Rev. Lett. **68**, 1943 (1992).

[30] P. Carra, B.T. Thole, M. Altarelli, and X. Wang, Phys. Rev. Lett. **70**, 694 (1993).

[31] Y. Teramura, A. Tanaka, and T. Jo, J. Phys. Soc. Jpn. **65**, 1053 (1996).

[32] J. Kanamori, J. Phys. Chem. Solids **10**, 18 (1959).

[33] C. Piamonteze, P. Miedema and Frank M.F. Groot, Phys. Rev. B **80**, 184410 (2009).

[34] J. Rubio-Zuazo, A. Chainani, M. Taguchi, D. Malterre, A. Serrano, and G.R. Castro, Phys. Rev. B **97**, 235148 (2018).

[35] S.W. Han, J.-S. Kang, S.S. Lee, G. Kim, S.J. Kim, C.S. Kim, J.-Y. Kim, H. J. Shin, K.H. Kim, J.I. Jeong, B.-G. Park, J.-H. Park and B.I. Min, J. Phys.: Condens. Matter **18**, 7413 (2006).